# Electronic structure investigation of CeB$_6$ by means of soft X-ray scattering

M. Magnuson[*], S. M. Butorin, J.-H. Guo, A. Agui[**] and J. Nordgren
*Department of Physics, Uppsala University, P. O. Box 530, S-751 21 Uppsala, Sweden*

H. Ogasawara and A. Kotani
*Institute for Solid State Physics, University of Tokyo, Kashiwanoha, Kashiwa, Chiba, 277-8581, Japan*

T. Takahashi and S. Kunii
*Department of Physics, Tohoku University, Sendai 980-77, Japan*

**Abstract**

The electronic structure of the heavy fermion compound CeB$_6$ is probed by resonant inelastic soft X-ray scattering using photon energies across the Ce $3d$ and $4d$ absorption edges. The hybridization between the localized $4f$ orbitals and the delocalized valence-band states is studied by identifying the different spectral contributions from inelastic Raman scattering and normal fluorescence. Pronounced energy-loss structures are observed below the elastic peak at both the $3d$ and $4d$ thresholds. The origin and character of the inelastic scattering structures are discussed in terms of charge-transfer excitations in connection to the dipole allowed transitions with $4f$ character. Calculations within the single impurity Anderson model with full multiplet effects are found to yield consistent spectral functions to the experimental data.

## 1 Introduction

The electronic structures of Ce heavy fermion compounds has attracted much attention both from an experimental and theoretical point of view since they show a variety of unusual and interesting macroscopic properties. In particular, much interest has been focused on the $4f$ narrow-band occupancy, and the role of hybridization with the conduction band states which strongly affects the physical properties [1,2]. It remains a controversy if the localized Ce $4f$ states are best theoretically described with the Single Impurity Anderson Model (SIAM) [3,4] or with density functional theory [5].

Experimental methods such as X-ray absorption spectroscopy (XAS) and resonant photoemission spectroscpy (RPES) of the valence band and core levels have previously proved to be powerful tools for investigating the electronic states, in particular, close to the Fermi level [6,7,8]. Recent improvements in the energy resolution of valence-band photoemission spectra with excitation by high-brilliance undulator radiation allow detailed observations of the tail of the so-called Kondo resonance as a very narrow peak in the vicinity of the Fermi level of rather hybridized compounds [9]. However, the predominant surface sensitivity constitutes a serious inherent problem with the use of electron spectroscopies for these kind of studies. Since the $4f$ states in the Kondo resonance in the surface are very different from those in the bulk, the SIAM approach for RPES has been a subject of controversy [10,11,12,13].

To further investigate the electronic structures and the bulk properties of heavy fermion materials, we have chosen to study the well known hexaboride CeB$_6$ compound which is known to exhibit dense Kondo properties. The present measurements were performed using high-resolution resonant inelastic X-ray scattering (RIXS) spectroscopy with selective excitation energies at the Ce $3d$ and $4d$ thresholds. A general advantage of using the RIXS technique is the bulk sensitivity which makes it possible to avoid surface contributions. With the scattering process, the radiative transitions probe the ground and low-lying excited states [14]. Recent valence-band RPES measurements at the $3d$ and $4d$ thresholds of $CeB_6$ show two main resonating peak structures: a narrow Kondo peak at the Fermi edge, and a broad structure at 2.5 eV binding energy [15, 16]. According to SIAM calculations including two configurations, these peaks were assigned to the $4f^1$ and $4f^0$ final states, respectively.

In general, the identification of the energy positions of the $4f$ energy loss structures using the RIXS technique in different Ce compounds is essential for understanding the properties of heavy fermion systems. Although the final states are different in RIXS and RPES, it is useful to compare the energy positions of the peak structures and the validity of the SIAM in both cases for deriving more accurate values of the model parameters. When the excitation energy is tuned on and above the thresholds, the RIXS spectra of $CeB_6$ are found to exhibit interesting resonance behaviors with pronounced energy loss structures below the strong elastic peak. The origin and $4f$ character of the loss structures are discussed in connection to the hybridization with the delocalized states. The experimental RIXS spectra of $CeB_6$ are interpreted with the results of state-of-the-art SIAM calculations. Using the same set of SIAM parameters, the RIXS spectra and the XAS spectra provide complementary information. Due to the hybridization, the initial, intermediate and final states of the RIXS process are best described as mixtures of all three $4f^0$, $4f^1$ and $4f^2$ configurations.

## 2 Experimental Details

The measurements at the Ce $3d$ edge were performed at beamline BW3 at HASYLAB, Hamburg, using a modified SX700 monochromator [17]. An XAS spectrum at the Ce $3d$ edge was obtained in total electron yield (TEY) by measuring the sample drain current. The Ce $3d$ RIXS spectra were recorded using a high-resolution grazing-incidence grating spectrometer with a two-dimensional position-sensitive detector [18]. During the XAS and RIXS measurements at the Ce $3d$ edge, the resolutions of the beamline monochromator were about 0.6 eV and 2.0 eV, respectively. The X-ray emission spectrometer had a resolution of about 1.5 eV.

Measurements at the Ce $4d$ threshold were carried out at beamline 7.0 at the Advanced Light Source at the Lawrence Berkeley National Laboratory. The beamline comprises a 89-pole, 5 cm period undulator and a spherical-grating monochromator [19]. An XAS spectrum at the $4d$ threshold region was also obtained in TEY by measuring the sample drain current. During the XAS and RIXS measurements at the $4d$ edge, the resolution of the monochromator of the beamline was $\sim 0.1$ eV. The X-ray emission spectrometer had a resolution better than 0.2 eV.

All the measurements at the Ce $3d$ and $4d$ thresholds were performed at room temperature with a base pressure lower than $5 \times 10^{-9}$ Torr. In order to minimize self-absorption effects, the sample crystal was oriented so that the photons were incident at an angle of about $20^o$ with respect to the sample surface. The emitted photons were recorded at an angle, perpendicular to the direction of the incident photons, with the polarization vector parallel to the horizontal scattering plane. The single crystal of $CeB_6$ was grown by the floating-zone method [20].

## 3 Calculational Details

The Ce $4f \rightarrow 3d$ and Ce $4f \rightarrow 4d$ RIXS spectra of $CeB_6$ were calculated as a coherent second-order optical process including interference effects using the Kramers-Heisenberg formula [21]. The Slater integrals, describing $4f$–$4f$, $4f$–$3d$ and $4f$–$4d$ Coulomb and exchange interactions, and spin-orbit constants were obtained by the Hartree-Fock method with relativistic corrections [22]. The reduction ratios of the Slater integrals were $F^k(4f4f)$ 80%, $F^k(3d4f)$ 80%, $G^k(3d4f)$ 80%, $F^k(4d4f)$ 75%, and $G^k(4d4f)$ 66%. Three different configurations were considered: $4f^0$, $4f^1$, $4f^2$ for the initial and final states, and $d^9 4f^1$, $d^9 4f^2$, $d^9 4f^3$ for the intermediate states. The weights of the $f^0$, $f^1$ and $f^2$ configurations in the ground state were about 1%, 97% and 2%, respectively. The effect of electron-hole pair excitations in the conduction band and the crystal-field effects were disregarded for simplicity. The calculations were made at zero Kelvin, so that the ground state was in the Kondo singlet $^1S_0$ bound state due to the effect of hybridization. The scattering angle was fixed to 90 degrees and calculations were made for two different geometries where the scattering plane was either parallel ("depolarized geometry") or perpendicular ("polarized geometry") to the polarization vector of the incident photons.

The SIAM [3] with full multiplet effects was used to describe the system. The parameters were chosen to reproduce the experiment as follows: the $4f$ level (with respect to the Fermi level $\varepsilon_F$) $\varepsilon_f - \varepsilon_F = -2.5$, the occupied conduction bandwidth $W = 3.0$, the hybridization strength between the $4f$ and conduction band states $V = 0.15$, the Coulomb interaction between the $4f$ electrons $U_{ff} = 6.5$, and that between $4f$ and $3d$ core electrons $U_{fc}(3d) = 8.5$ (in units of eV). The Coulomb interaction between the $4f$ and $4d$ core electrons $U_{fc}(4d)$ was taken to be 80% of $U_{fc}(3d)$. The effect of the configuration-dependent hybridization was taken into account with the two reduction factors $R_c = 0.8$ and $R_v = 0.9$ [23]. The parameters used in the calculations are summarized in Table I. Similar SIAM parameters have previously been used for valence and core level photoemission [15].

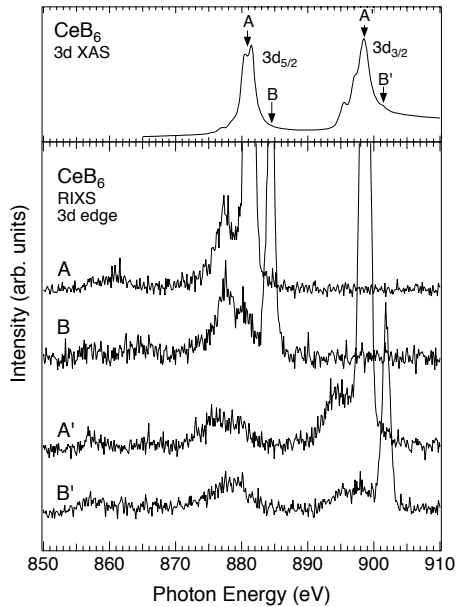

**Figure 1:** An XAS spectrum (upper panel) of CeB$_6$ measured at the $3d$ edge. A set of RIXS spectra (lower panel) for different photon energies around the $3d$ edge of CeB$_6$ on a photon energy scale. The measurements were made at room temperature at the following excitation energies for the RIXS spectra: 881.2 eV, 884.8 eV, 898.5 eV and 901.8 eV.

## 4 Results and Discussion

Figure 1 shows a set of RIXS spectra of CeB$_6$ recorded at different excitation energies near the Ce $3d_{5/2}$, $3d_{3/2}$ thresholds. In the top panel, an XAS spectrum is shown where the excitation energies for the RIXS spectra are indicated by the arrows aimed at the main peaks A(A') and weak satellites B(B') above the thresholds, respectively. The final states of the XAS spectrum are the same as the intermediate states in the RIXS process. The X-ray emission spectra in the lower panel basically consist of three parts: the elastic peak at an energy position equivalent to the excitation energy, the inelastic scattering contribution, also referred to as Raman scattering, and the normal fluorescence. Since the weight of the $f^1$ configuration dominates in the ground state of CeB$_6$, the normal fluorescence intensity is mainly due to decays through the $3d^9 4f^1 \rightarrow 3d^{10} 4f^0$ transitions, where the final states have one electron less than in the ground state. The elastic peak, also referred to as Rayleigh scattering, is due to the $3d^9 4f^2 \rightarrow 3d^{10} 4f^1$ transitions back to the ground state. In order to minimize the intensity of the elastic peak, which is very strong due to the localization of the $4f$ electrons, the measurements were made in the depolarized geometry. The relative intensity of the strong elastic peak as well as the weaker inelastic scattering spectral contributions exhibit strong intensity variations with varying excitation energy.

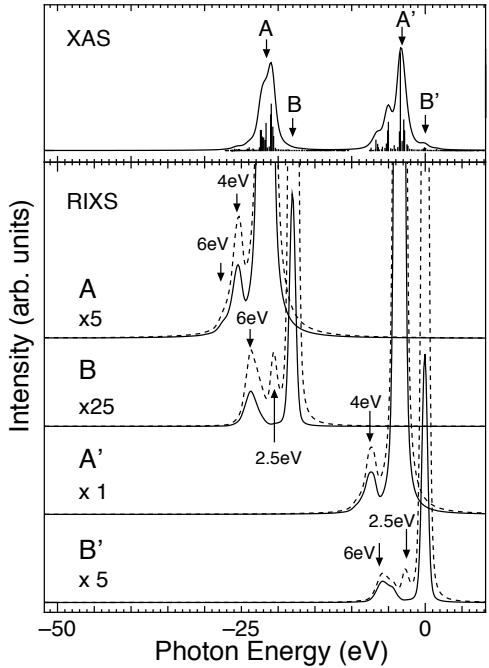

**Figure 2:** A calculated isotropic XAS spectrum (upper panel) and RIXS spectra (lower panel) for CeB$_6$ at the $3d$ edge. The XAS spectrum was convoluted with a Lorentzian of $\Gamma$=0.4 eV (HWHM) and Gaussian of $\sigma$=0.3 eV (FWHM). The RIXS spectra were convoluted with a Lorentzian of $\Gamma$=0.1 eV (HWHM) and a Gaussian of $\sigma$=0.4 eV(FWHM).

In the topmost RIXS spectrum, excited at the $3d_{5/2}$ peak maximum at A, an inelastic scattering feature is observed at about 4 eV energy loss below the elastic peak. In the next spectrum, excited at the weak satellite above the absorption threshold at B, the 4 eV peak is still visible. However, another peak at about 6-7 eV energy loss is now stronger. It should be noted that the 6-7 eV energy loss feature is not related to a normal $3d_{5/2}$ fluorescence line. In spectra A' and B', the inelastic scattering features are still found at about 4.0 eV and 6-7 eV below the elastic peak and a normal $3d_{5/2}$ fluorescence line now shows up as a broad structure between 875-880 eV photon energy. Similar low-energy loss structures as those observed at 4 eV and 6-7 eV, have previously been assigned to have charge-transfer origin showing a resonant behaviour at the thresholds in RIXS spectra of CeO$_2$ and UO$_3$ [14] and in electron energy loss spectra of metallic Ce [24,25]. The charge-transfer excitations occur as a result of electron hopping from delocalized states to a localized state. Weak emission features are also observed in the energy region between 850 and 870 eV due to $5p \rightarrow 3d$ transitions.

Figure 2 shows the results of SIAM calculations of the $3d$ XAS and RIXS spectra of CeB$_6$. The letters A(A') and B(B') denote the same excitation energies as in Fig. 1. The XAS spectrum in the upper panel is made up of the hybridized $3d^9 4f^1$, $3d^9 4f^2$ and $3d^9 4f^3$ manifolds, where the $3d^9 4f^2$ configuration dominates. The RIXS calculations, shown in the lower panel, are in good agreement with the experimental data in Fig. 1 although contributions from normal fluorescence are not included. Three different kinds of low-energy excited loss structures are observed at 2.5 eV, 4 eV and 6 eV below the elastic peak. The full line shows the results in the depolarized geometry i.e., in the same geometry as the experimental results in Fig. 1. In this geometry, the recombination back to the $^1S_0$ ground state is forbidden from geometrical selection rules [26], so that the true elastic line is forbidden. However, $4f^1$ final states with other symmetries give rise to a relatively strong quasi-elastic peak.

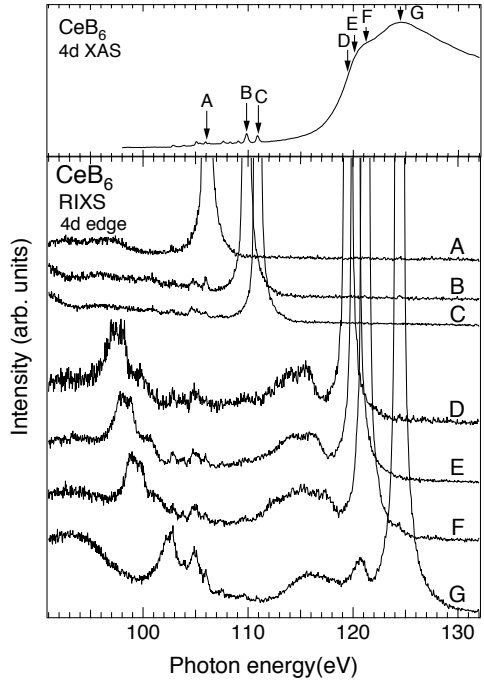

**Figure 3:** An XAS spectrum (top) of $CeB_6$ measured at the $4d$ edge. A set of RIXS spectra (bottom) for different photon energies around the $4d$ edge of $CeB_6$ on a photon energy scale. The measurements were made at room temperature at the following excitation energies for the RIXS spectra: 106.0 eV, 109.9 eV, 110.9 eV, 119.5 eV, 120.2 eV, 121.2 eV and 124.5 eV.

The 4 eV and 6 eV inelastic scattering structures which stay at constant energy loss can be identified in all the simulated RIXS spectra. These structures are almost entirely due to the $4f^2$ final states ($f^1 \rightarrow f^2$ charge transfer excitations as a result of hybridization), and it is found that the 4 eV peak corresponds mainly to triplet final state terms ($^3F_{2,3,4}$ and $^3H_{4,5,6}$), while the 6 eV peak is mainly due to singlet final state terms ($^1I_6$ and $^1D_2$). The lower energy peak (4 eV) is resonantly enhanced by the intermediate states of the main XAS peaks A(A'), while the higher energy peak (6 eV) is resonantly enhanced by the intermediate states of the XAS satellites B(B'). The dashed lines show the results in the polarized geometry i.e., when the scattering plane is perpendicular to the polarization vector of the incoming photons. In this configuration, a low energy excited peak structure of $4f^0$ character appears at 2.5 eV energy loss from the elastic peak. The $f^0$ final states occur as an indication of the singlet (Kondo) ground state. The transitions to the $f^0$ final states are forbidden in the depolarized geometry due to the $^1S_0$ symmetry of the Kondo ground state.

Figure 3 (top panel) shows an XAS spectrum of $CeB_6$ recorded at the Ce $4d$ threshold. The large broad structure above threshold in the energy region between 120-130 eV, commonly referred to as the giant absorption region, has been observed in many Ce compounds [27]. The series of small sharp peaks in the pre-threshold region around 102-112 eV have been interpreted as due to transitions to various multiplet states of the excited $4d^9 4f^2$ configuration [28]. The lower panel of Fig. 3 shows a set of RIXS spectra recorded across the Ce $4d$ threshold of $CeB_6$. The spectra are plotted on an emission photon energy scale and were measured at excitation energies denoted by the letters A-G in the XAS spectrum from 106.0 eV up to 124.5 eV. Pronounced inelastic scattering structures are observed below the strong elastic peaks. The X-ray emission spectra at the $4d$ edge can basically be divided into three parts in the same way as for those measured at the $3d$ edge. However, the prominent peak structures observed at about 95-105 eV photon energy are attributed to Ce $5p_{3/2,1/2}$→4$d$ transitions previously observed in RPES [15]. In the energy region above the 5$p$ peaks, several smaller sharp peaks are observed in the spectra. These fine structures are related to those in the pre-threshold region of the XAS spectrum.

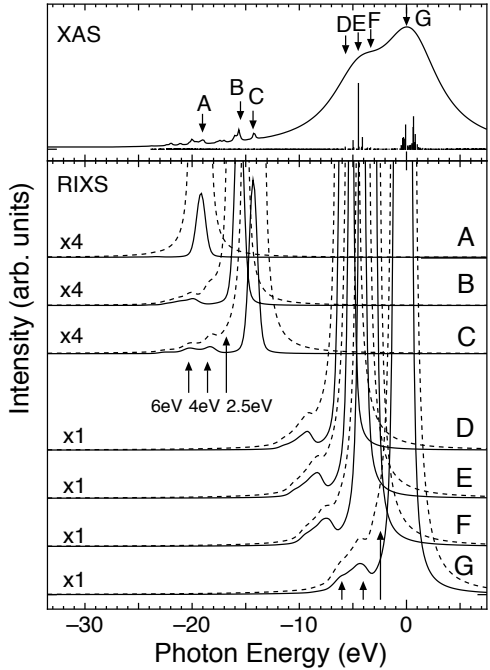

**Figure 4:** A calculated isotropic XAS spectrum (upper panel) and RIXS spectra (lower panel) for CeB$_6$ at the 4$d$-edge. The XAS spectrum was convoluted with a Lorentzian of Γ=0.4 eV (HWHM) and a Gaussian of σ=0.3 eV (FWHM). The RIXS spectra were convoluted with a Lorentzian of Γ=0.1 eV (HWHM) and a Gaussian of σ=0.2 eV (FWHM).

A comparison to previous RPES measurements of CeB$_6$ [29], shows that the 4$f$ and 5$p$ intensities has a resonance maximum at about 121 eV. The major parts of the resonances are thus expected to appear at the shoulder of the broad absorption structure of the giant band. At somewhat higher photon energies (127 eV), the RPES spectra are instead dominated by the Auger decay, corresponding to normal fluorescence in the radiative channel. In Fig. 3, the normal fluorescence contribution appears below the pronounced 4 eV peak as a broad structure at ~ 9 eV loss energy in spectrum G. A minor contribution to the intensity distribution of the energy loss features in spectra D-F may also be attributed to normal fluorescence, although the major part of the intensity is related to inelastic scattering.

Figure 4 shows the results of SIAM calculations of the 4$d$ XAS and RIXS processes for CeB$_6$. The letters A-G denote the same excitation energies as in Fig. 3. The results are in good agreement with the experiment with the exception of the 5$p$ emission lines and the normal fluorescence which are not included in the calculation. The charge-transfer satellites to the $f^2$ final states at 4 eV and 6 eV below the elastic line behave in the simulation in a similar way as at the 3$d$ edge. The 4 eV feature in the RIXS spectra (D-G) is resonantly enhanced at the threshold energy i.e., at the shoulder of the giant absorption region for excitation energies around 118-125 eV, while the relatively weak spectral weight of the 6 eV feature has a maximum at G i.e., at 124.5 eV excitation energy. The $f^0$ final states at 2.5 eV in the polarized geometry (dashed lines) appear as a small shoulder at the low-energy side of the strong elastic peak.

**Table 1:** Parameter values used in the Anderson impurity model calculations. $W$ is the width of the occupied conduction band, which is discretized approximately by $N$ levels. $\varepsilon_f$-$\varepsilon_F$ is the 4$f$ level measured from the Fermi level, $U_{ff}$ is the Coulomb interaction between the 4$f$ electrons and $U_{fc}(3d)$ between the 4$f$ and 3$d$ electrons. $V$ is the hybridization strength and the parameters $R_c$ and $R_v$ are configuration-dependent reduction factors. The values of $W$, $\varepsilon_f$-$\varepsilon_F$, $U_{ff}$, $U_{fc}(3d)$ and $V$ are in units of eV.

| $N$ | $W$ | $\varepsilon_f$-$\varepsilon_F$ | $U_{ff}$ | $U_{fc}(3d)$ | $V$ | $R_c$ | $R_v$ |
|---|---|---|---|---|---|---|---|
| 4 | 3.0 | −2.5 | 6.5 | 8.5 | 0.15 | 0.8 | 0.9 |

The interpretation of the relative peak positions of the low-energy loss structures in the RIXS spectra are consistent with the results of RPES which show a narrow $f^1$ Kondo peak at the Fermi edge and a broad $f^0$ structure at 2.5 eV binding energy. Due to the configuration interaction between the Ce $4f$ states, the $f^1$ energy loss structure is located at 0 eV and the $f^0$ structure is predicted to show up at 2.5 eV energy loss in the RIXS spectra when measurements are made in the polarized geometry for temperatures close to or below the Kondo temperature $T_K$. For higher temperatures, the spectra will appear similar to those of a normal $Ce^{3+}$ system, even though the true ground state is the $^1S_0$ Kondo singlet. In the depolarized geometry where the $f^0$ final states are forbidden, the spectra look similar to a normal $Ce^{3+}$ system, even at zero Kelvin. It is interesting to notice that the $f^2$ final states observed at 4 eV and 6 eV energy loss in the RIXS spectra cannot be observed in valence band RPES but instead appears at 2 eV in inverse photoemission[7]. The difference in the energy position of the $f^2$ peak between the different spectroscopical techniques is a result of probing the systems with different number of electrons in the final states in comparison to the ground state. The RIXS technique is thus shown here to be very useful for detecting the $f^1 \rightarrow f^2$ charge transfer excitations. Since the RIXS process is charge neutral, it is a very powerful tool for investigating the ground and low-energy excited states of systems where charge-transfer excitations are important. In particular, future utilization of angular polarization dependent measurements of the inelastic scattering features below the Kondo temperature will allow interesting studies of the $f^1 \rightarrow f^0$ charge transfer excitations.

# 5 Conclusions

The electronic structure of $CeB_6$ has been measured at the Ce $3d$ and $4d$ absorption thresholds by resonant inelastic soft-X-ray scattering. Due to the bulk sensitivity, the resonant inelastic soft X-ray scattering technique allows probing the ground and low-lying $4f$ states without surface effects. By changing the incoming photon energy, the contribution from inelastic scattering is distinguished from the normal fluorescence. Pronounced energy loss structures below the strong elastic peaks are identified as due to charge-transfer excitations to the $4f^2$ state as a result of configurational mixing in the ground and core-excited states. Comparisons to peak structures in model calculations within the single impurity Anderson model and valence band resonant photoemission are consistent with the present experimental results.

# 6 Acknowledgments

This work was supported by the Swedish Natural Science Research Council (NFR), the Göran Gustavsson Foundation for Research in Natural Sciences and Medicine. ALS is supported by the U.S. Department of Energy, under contract No. DE-AC03-76SF00098. T. Takahashi and A. Kotani would like to acknowledge a Grant-in-Aid from the Ministry of Education, Science, Culture and Sports in Japan. The computation in this work has been done using the facilities of the Supercomputer Center, ISSP, University of Tokyo.

# References

[*]  Present address: Université Pierre et Marie Curie (Paris VI), Laboratoire de Chimie Physique - Matière et Rayonnement (UMR 7614), 11 rue P. et M. Curie, F-75231 Paris Cedex 05, France.

[**] Present address: JAERI, SPring-8, 1-1-1 Kouto, Mikazuki, Sayo, Hyogo 679-5148, Japan.

[1]  See e.g., J. W. Allen, S. J. Oh, O. Gunnarsson, K. Schönhammer, M. B. Maple, M. S. Torikachvili and I. Lindau; Advances in Physics; **35**, 275 (1986).

[2]  D. Malterre, M. Grioni, and Y. Baer; Adv. Phys. **45**, 299 (1996).

[3]  P. W. Anderson, Phys. Rev. B **124**, 41 (1961).

[4]  A. Kotani; J. Electr. Spec. & Relat. Phenom., *in press*; **100**, 75 (1999); **92**, 171 (1998); **78**, 7 (1996).

[5]  M. B. Suvasini, G. Y. Guo, W. M. Temmerman, G. A. Gehring and M. Biasini; J. Phys. Condens. Matter **8**, 7105 (1996).

[6]  S. Kimura, T. Nanba, S. Kunii and T. Kasuya; Phys. Rev. B **50**, 1406 (1994).

[7]  N. Shino, S. Suga, S. Imada, Y. Saitoh, H. Yamada, T. Nanba, S. Kimura and S. Kunii; J. Phys. Soc. Jpn. **64**, 2980 (1995).

[8]  T. Takahashi, T. Morimoto, T. Yokoya, S. Kunii, T. Komatsubara and O. Sakai; Phys. Rev. B **52**, 9140 (1995).

[9]  A. Sekiyama, T. Iwasaki, K. Matsuda, Y. Saitoh, Y. Onuki and S. Suga; Nature **403**, 396 (2000).

[10] J. J. Joyce, A. J. Arko, J. Lawrence, P. C. Canfield, Z. Fisk, R. J. Bartlett and J. D. Thompson; Phys. Rev. Lett. **68**, 236 (1992).

[11] L. Duo, S. De Rossi, P. Vavassori, F. Ciccacci, G. L. Olcese, G. Chiaia and I. Lindau; Phys. Rev. B **54**, R17363 (1996).

[12] A. J. Arko and J. J. Joyce; Phys. Rev. Lett. **81**, 1348 (1998).

[13] M. Garnier, K. Breuer, D. Purdie, M. Hengsberger and Y. Baer; Phys. Rev. Lett. **78**, 4127 (1997).

[14] S. M. Butorin, D. C. Mancini, J.-H. Guo, N. Wassdahl, J. Nordgren, M. Nakazawa, S. Tanaka, T. Uozumi, A. Kotani, Y. Ma, K. E. Myano, B. A. Karlin, D. K. Shuh; Phys. Rev. Lett. **77**, 574 (1996).

[15] A. Kakizaki, A. Harasawa, T. Ishii, T. Kashiwakura A. Kamata and S. Kunii; J. Phys. Soc. Jap. **64**, 302 (1995).

[16] G. Chiaia, O. Tjernberg, L. Duo, S. De Rossi, P. Vavassori, I Lindau, T. Takahashi, S. Kunii, T. Komatsubara, D. Cocco, S. Lizzit and G. Paolucci; Phys. Rev. B **55**, 9 207 (1997).

[17] T. Möller; *Synchrotron Radiat. News.* **6**, 16 (1993).

[18] J. Nordgren and R. Nyholm, Nucl. Instr. Methods A**246**, 242 (1986); J. Nordgren, G. Bray, S. Cramm, R. Nyholm, J.-E. Rubensson, and N. Wassdahl, Rev. Sci. Instrum. **60**, 1690 (1989).

[19] T. Warwick, P. Heimann, D. Mossessian, W. McKinney and H. Padmore; Rev. Sci. Instr. **66**, 2037 (1995).

[20] N. Sato, S. Kunii, I. Oguro, T. Komatsubara, T. Kasuya J. Phys. Soc. Jpn., **53**, 3967 (1984).

[21] H. H. Kramers and W. Heisenberg, Zeits. f. Phys. **31**, 681 (1925).

[22] R. D. Cowan, *The Theory of Atomic Structure and Spectra* (University of California Press, Berkeley, 1981).

[23] M. Nakazawa, S. Tanaka, T. Uozumi and A. Kotani, J. Phys. Soc. Jpn. **65**, 2303 (1966).

[24] J. Bloch, N. Shamir, M. H. Mintz and U. Atzmony; Phys. Rev. B **30**, 2462 (1984).

[25] F. P. Netzer, G. Strasser, G. Rosina and J. A. D. Matthew; Surface Science **152/153**, 757

(1985).

[26]   M. Nakazawa, H. Ogasawara, A. Kotani and P. Lagarde, J. Phys. Soc. Jpn., **67**, 323 (1998).

[27]   D. Wieliczka, J. H. Weaver, D. H. Lynch and C. G. Olson; Phys. Rev. B **26**, 7056 (1982).

[28]   K. Ichikawa, A. Nisawa and K. Tsutsumi; Phys. Rev. B **34**, 6690 (1986).

[29]   H. Sugawara, A. Kakizaki, I. Nagakura, T. Ishii, T. Komatsubara and T. Kasuya;
       J. Phys. Soc. Jpn. **51**, 915 (1982).